\documentclass[english,reprint,aps,amsmath,amssymb,prl,superscriptaddress,floatfix]{revtex4-1}
\usepackage[T1]{fontenc}
\usepackage{color}
\usepackage{bm}
\usepackage{varwidth}
\usepackage{amsmath}
\usepackage{amssymb}
\usepackage{graphicx}

\makeatletter

\providecommand{\tabularnewline}{\\}
\newenvironment{cellvarwidth}[1][t]
    {\begin{varwidth}[#1]{\linewidth}}
    {\@finalstrut\@arstrutbox\end{varwidth}}

\usepackage{dcolumn}
\usepackage{bm}
\usepackage{mathrsfs}
\usepackage[colorlinks,linkcolor=red,anchorcolor=blue,citecolor=green]{hyperref}
\allowdisplaybreaks[4]

\usepackage{amssymb}
\makeatletter

\newcommand{\Rmnum}[1]{\expandafter\@slowromancap\romannumeral #1@}
\makeatother

\makeatother

\usepackage{babel}
\begin{document}
\title{Phonon Dichroisms Revealing Unusual Electronic Quantum Geometry }
\author{Ding Li}
\email{Both authors contribute equally to this work.}

\affiliation{Anhui Provincial Key Laboratory of Low-Energy Quantum Materials and
Devices, High Magnetic Field Laboratory, HFIPS, Chinese Academy of
Sciences, Hefei, Anhui 230031, China}
\affiliation{Department of Physics, University of Science and Technology of China,
Hefei 230026, P.R. China}
\author{Guoao Yang}
\email{Both authors contribute equally to this work.}

\affiliation{School of Physics and Optoelectronics Engineering, Anhui University,
Hefei, Anhui Province 230601, P.R. China}
\affiliation{Anhui Provincial Key Laboratory of Low-Energy Quantum Materials and
Devices, High Magnetic Field Laboratory, HFIPS, Chinese Academy of
Sciences, Hefei, Anhui 230031, China}
\author{Tao Qin}
\email{taoqin@ahu.edu.cn}

\affiliation{School of Physics and Optoelectronics Engineering, Anhui University,
Hefei, Anhui Province 230601, P.R. China}
\author{Jianhui Zhou}
\email{jhzhou@hmfl.ac.cn}

\affiliation{Anhui Provincial Key Laboratory of Low-Energy Quantum Materials and
Devices, High Magnetic Field Laboratory, HFIPS, Chinese Academy of
Sciences, Hefei, Anhui 230031, China}
\author{Yugui Yao}
\email{ygyao@bit.edu.cn}

\affiliation{Centre for Quantum Physics, Key Laboratory of Advanced Optoelectronic
Quantum Architecture and Measurement, School of Physics, Beijing Institute
of Technology, Beijing 100081, China}
\affiliation{Beijing Key Lab of Nanophotonics and Ultrafine Optoelectronic Systems,
School of Physics, Beijing Institute of Technology, Beijing 100081,
China}
\begin{abstract}
The quantum geometry tensor, intrinsic geometric characteristics of
electronic states, plays a crucial role in the various nontrivial
electromagnetic phenomena in quantum materials. Here, we reveal that
quantum geometry significantly modifies phonon dichroisms through
electron-phonon interactions in solids that break time-reversal and
spatial inversion symmetries. Specifically, the circular phonon dichroism
is primarily dominated by the heat magnetic moments, while the linear
phonon dichroism depends on the heat Drude weight, a thermal analog
of band Drude weight. Furthermore, we establish the $f$-sum rule
for the heat magnetic moment that facilitates its experimental detections.
We demonstrate our key findings in an archetypal model system: ferromagnetic
two-dimensional electron gases with Rashba spin-orbit coupling. Our
work uncovers the quantum-geometric origin of common phonon dichroisms
and predicts the detectable signature of the heat magnetic moment
of electrons in solids. 
\end{abstract}
\maketitle
\textit{\textcolor{black}{Introduction.-{}-}}\textcolor{black}{The
}familiar\textcolor{black}{{} quantum geometry tensor characterizes
the }fundamental\textcolor{black}{{} geometric properties of nearby
quantum states }in Hilbert space\textcolor{black}{, whose imaginary
part and real part correspond to the Berry curvature and quantum metric,
respectively \citep{Provost1980CMP,Wilczek1989gp,Torma2023PRL}. The
Berry curvature modifies the electronic properties through an anomalous
velocity, which has played a vital role in understanding various dissipation-less
quantum effects \citep{XiaoD2010RMP,Nagaosa2010RMP} and the emergence
of the topological phases of matter \citep{Hasan2010RMP,QiXL2011RMP,Armitage2018RMP}.
Recently, the quantum metric has also manifested in a plenty of fundamental
phenomena in quantum materials \citep{MaYQ2010PRB,LiangL2017PRB1,LiangL2017PRB,Sengupta2020PRL,HuLH2021PRL,RenYF2021PRL,WangZ2021PRL,Torma2022nrp,Chen2024PRL,ZhangJL2024PRL},
such as the nonlinear Hall effect \citep{GaoY2014PRL,Sodemann2015PRL,GaoAY2023Science}.
However, the quantum geometry effect in the electron-collective excitation
(such as phonon, plasmon, magnon) coupled systems in conducting crystals
remains largely elusive \citep{YuJ2024NP,HuJM2024}. }

Controllable symmetry breaking of crystal primarily determines the
essential behaviors of \textcolor{black}{quantum geometry} tensor
and the resulting quantum phenomena \citep{XiaoD2010RMP,Nagaosa2010RMP}.
For example, the nontrivial Berry curvature in the momentum space
requires the breaking of either time reversal or spatial inversion
symmetry. The Berry curvature leads to the self-rotation of wave packet
of Bloch electrons which is accompanied with orbital magnetic moment
\citep{XiaoD2005PRL,Thonhauser2005PRL} and heat magnetic moment (HMM)
of electrons \citep{QinT2011PRL}. Remarkably, the orbital magnetic
moment is closely related to the dynamical Hall responses and can
be effectively probed by the magnetic circular dichroism \citep{Smith1976PRB,Thole1992PRL,Souza2008PRB,YaoW2008PRB,Kang2025NP,Bac2025PRL,Kim2025Science}.
Nevertheless, the manifestation, detection and manipulation of HMM
of electrons are still challenging, a prime reason is the lack of
independent and easily accessible quantum phenomena dominated by the
HMM of electrons. Here we focus on the key roles of quantum geometry
in phonon dichroisms (the differential absorption of linearly or circularly
polarized phonons) through the electron-phonon coupling, allowing
the current-free, contactless and acoustic detection of HMM of electrons. 

In this work, we uncover the intimate connection between unusual electronic
quantum geometry and phonon dichroism. Specifically, the circular
phonon dichroism is predominately driven by the HMMs of electrons,
while the linear phonon dichroism originates from the heat Drude weight.
We derive the $f$-sum rule of HMM to facilitate experimental detections,
in analogy to the orbital magnetization. Moreover, we demonstrate
our key findings in the representative ferromagnetic two-dimensional
electron gases (2DEG) with Rashba spin-orbit coupling (SOC) and brief
its experimental detections. 

\textit{\textcolor{black}{Formalism of phonon dichroisms.-{}-}}We
start from the equation of motion for ions in metals:
\begin{align}
\omega^{2}u_{\alpha}\left(\boldsymbol{q}\right) & =\sum_{\beta}\left(\hbar^{2}\Phi_{\alpha\beta}\left(\boldsymbol{q}\right)+\hbar\chi_{\alpha\beta}\left(\boldsymbol{q},\omega\right)\right)u_{\beta}\left(\boldsymbol{q}\right),
\end{align}
where $\hbar\omega$ is the phonon energy, $u_{\alpha}(\boldsymbol{q})$
is the Fourier transform of the displacement field along the $\alpha$th
direction $u_{\alpha}\left(\boldsymbol{r}\right)$, and $\Phi_{\alpha\beta}(\boldsymbol{q})$
is the dynamic matrix. $\chi_{\alpha\beta}(\boldsymbol{q},\omega)$
is the retarded force-force response function \citep{LiuDH2017PRL}
\begin{figure}[h]
\includegraphics[width=8cm]{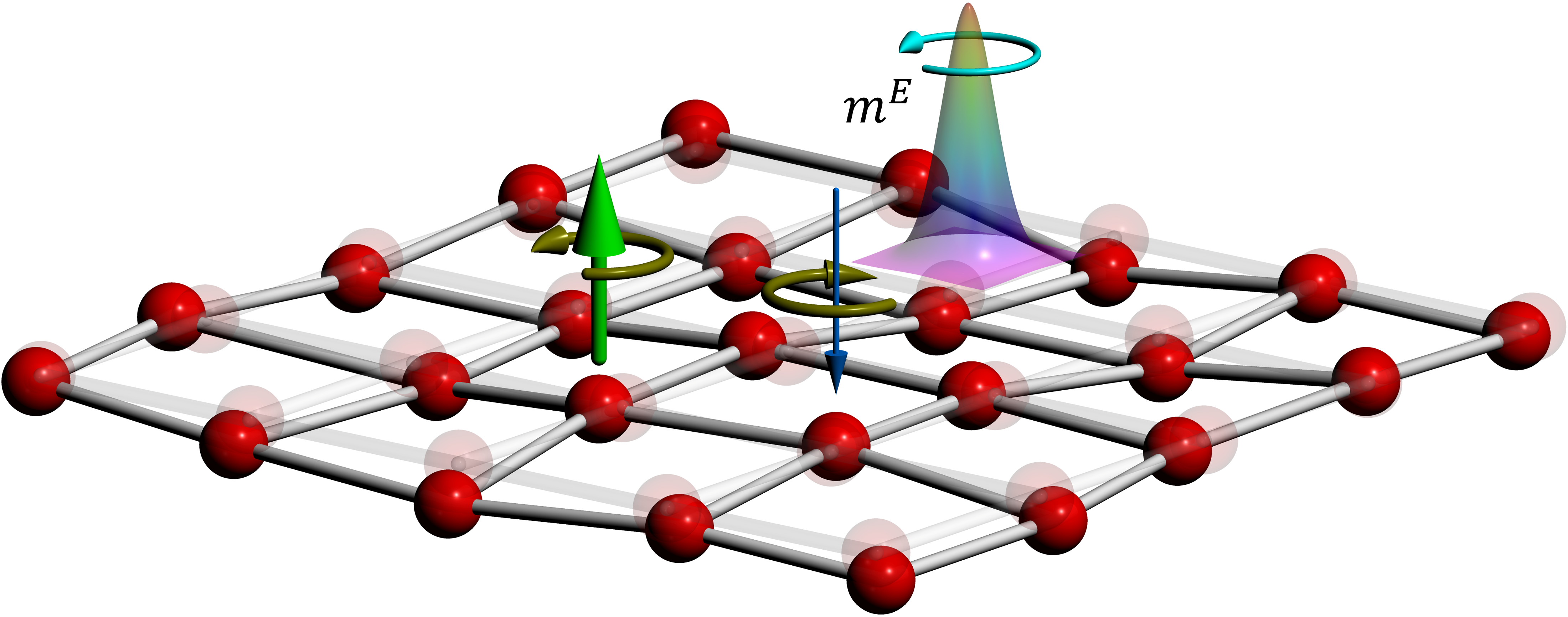}

\caption{\label{fig1} Schematic of the magnetic circular dichroism via electron-phonon
coupling due to the the heat magnetic moments ($\boldsymbol{m}^{E}$)
of the self-rotating electronic wave packet. The light and solid atoms
indicate the lattice vibration. The green up (blue down) arrow refers
to the absorptions of the right (left) circularly polarized phonons.
The sizes of arrows indicate different strength of absorptions.}
\end{figure}

\begin{align}
\chi_{\alpha\beta}(\boldsymbol{q},\omega) & =\sum_{n,n'}\int\frac{\hbar d^{2}\boldsymbol{k}}{\rho(2\pi)^{2}}F_{n'n}\left(\omega,\boldsymbol{k},\boldsymbol{q}\right)S_{nn^{\prime}}^{\alpha\beta}\left(\boldsymbol{k},\boldsymbol{q}\right),
\end{align}
where $\rho$ is the 2D mass density, the dynamical factor becomes
$F_{n^{\prime}n}(\omega,\boldsymbol{k},\boldsymbol{q})=\frac{f_{n,n^{\prime}}\left(\boldsymbol{k},\boldsymbol{k}^{\prime}\right)}{\hbar\omega+i\delta+\varepsilon_{nn^{\prime}}\left(\boldsymbol{k},\boldsymbol{k}+\boldsymbol{q}\right)}$
with the energy difference $\varepsilon_{mn}\left(\boldsymbol{k},\boldsymbol{k}^{\prime}\right)\equiv\varepsilon_{m}\left(\boldsymbol{k}\right)-\varepsilon_{n}\left(\boldsymbol{k}^{\prime}\right)$
and the distribution-function difference $f_{n,n^{\prime}}\left(\boldsymbol{k},\boldsymbol{k}^{\prime}\right)\equiv f_{n}\left(\boldsymbol{k}\right)-f_{n^{\prime}}\left(\boldsymbol{k}^{\prime}\right)$.
$S_{nn^{\prime}}^{\alpha\beta}(\boldsymbol{k},\boldsymbol{q})$ is
the product of the matrix elements of the force operator $\hat{T_{\alpha}}(\boldsymbol{q})=\text{\ensuremath{-\partial\hat{H}_{e-ph}/\partial u_{\boldsymbol{q}\alpha}}}$
\citep{SMs}
\begin{align}
S_{nn^{\prime}}^{\alpha\beta}(\boldsymbol{k},\boldsymbol{q}) & =\left\langle \psi_{n}(\boldsymbol{k})|\hat{T_{\alpha}}(-\boldsymbol{q})|\psi_{n'}(\boldsymbol{k}+\boldsymbol{q})\right\rangle \nonumber \\
 & \times\left\langle \psi_{n'}(\boldsymbol{k}+\boldsymbol{q})|\hat{T_{\beta}}(\boldsymbol{q})|\psi_{n}(\boldsymbol{k})\right\rangle ,
\end{align}
which encodes the intrinsic geometric information of quantum states
during the electron excitations by phonons, dubbed quantum geometric
factor. Here $n$ is the energy band index and $f_{n}(\boldsymbol{k})$
is the Fermi distribution function. $\varepsilon_{n}(\boldsymbol{k})$
denotes the energy dispersion of Bloch states $\left|\psi_{n}(\boldsymbol{k})\right\rangle $. 

In order to describe the dissipation of phonons, we introduce a quantity
$\gamma=\frac{i}{4\omega}(\chi-\chi^{\dagger})$ as 
\begin{equation}
\gamma=\left[\begin{array}{cc}
\gamma_{xx} & \gamma_{xy}\\
\gamma_{yx} & \gamma_{yy}
\end{array}\right]\label{gammaM}
\end{equation}
with $\gamma_{\alpha\beta}(\boldsymbol{q},\omega)=\frac{-1}{2\omega}\sum_{n,n'}\int\frac{\hbar d^{2}\boldsymbol{k}}{\rho(2\pi)^{2}}\mathrm{Im}[F_{n^{\prime}n}(\omega,\boldsymbol{k},\boldsymbol{q})]\times S_{nn^{\prime}}^{\alpha\beta}(\boldsymbol{k},\boldsymbol{q})$.
We define $\gamma_{D}=(\gamma_{xx}+\gamma_{yy})/2$ and $\gamma_{\bar{D}}=(\gamma_{xx}-\gamma_{yy})/2$,
which refer to the longitudinal absorptions, and furthermore, introduce
$\gamma^{A}=\mathrm{Im}[\gamma_{xy}]$, related to the anomalous Hall
absorption, and $\gamma^{\bar{A}}=\mathrm{Re}\left[\gamma_{xy}\right]$.
For the linearly polarized longitudinal and transverse phonons $\left(\omega_{l/t}=c_{l/t}\left|\boldsymbol{q}\right|\right)$,
the absorption coefficients are given by 
\begin{equation}
\gamma^{l/t}=\gamma_{D}\pm\mathrm{cos}2\phi_{q}\gamma_{\bar{D}}\pm\mathrm{sin}2\phi_{q}\gamma^{\bar{A}},\label{gamlt}
\end{equation}
with $\phi_{q}=\mathrm{tan^{-1}}(q_{y}/q_{x})$ \citep{ShanWY2022PRB},
where $\boldsymbol{q}$ is the phonon wave vector and $c_{l/t}$ is
the phonon velocity. For the left- and right-hand circularly polarized
phonons $\left(\frac{1}{2}\left(1,\pm i\right)^{T}\right)$, the absorption
coefficients are given by $\gamma^{L/R}=\gamma_{D}\mp\gamma^{A}$.
The difference between $\gamma^{l}$and $\gamma^{t}$ defines the
linear phonon dichroism, while the difference between $\gamma^{L}$
and $\gamma^{R}$ causes the circular phonon dichroism $(\gamma^{A}\equiv\left(\gamma^{R}-\gamma^{L}\right)/2)$
(see Fig.$\,$\ref{fig1} for the schematic of its detection).

\textit{\textcolor{black}{Quantum-geometric origin.-{}-}}In order
to uncover the clear quantum-geometric origin of phonon dichroisms,
we consider a generic two-band model 
\begin{equation}
H=d_{0}\left(\boldsymbol{k}\right)+\boldsymbol{d}\left(\boldsymbol{k}\right)\cdot\boldsymbol{\sigma},
\end{equation}
where the Pauli matrix $\sigma_{i}$ could denote the real spin and
pseudospin degree of freedom (such as lattice, orbit, layer) in the
context of graphene \citep{Neto2009RMP}, monolayer transition metal
dichalcogenides \citep{XiaoD2012PRL,CaoT2012NC}, the Rashba 2DEG
\citep{BychkovJETP1984,ZhouJH2015PRB}, Weyl semimetals \citep{WanXG2011PRB}
and topological surface states \citep{TseWK2010PRL}. Without loss
of generality, we assume that $d_{i}$ are the polynomial function
of $k_{x}$ and $k_{y}$. Before proceeding, we should brief the six
quantum geometric quantities (See Table. \ref{sixQGQ}). They can
be expressed in terms of the unit vector $\hat{\boldsymbol{d}}\equiv\boldsymbol{d}/\left|\boldsymbol{d}\right|$
with $\left|\boldsymbol{d}\right|=\sqrt{d_{1}^{2}+d_{2}^{2}+d_{3}^{2}}$:
\begin{align}
\mathcal{F}_{n\alpha\beta}\left(\boldsymbol{k}\right) & =\mathcal{G}_{n\alpha\beta}\left(\boldsymbol{k}\right)/2\left|\boldsymbol{d}\right|=\mathcal{H}_{n\alpha\beta}\left(\boldsymbol{k}\right)/4\left|\boldsymbol{d}\right|^{2},\label{FGH}
\end{align}
where the conventional quantum geometry tensor is given as $\mathcal{F}_{n\alpha\beta}\left(\boldsymbol{k}\right)=g_{n\alpha\beta}\left(\boldsymbol{k}\right)-\frac{i}{2}\Omega_{n\alpha\beta}\left(\boldsymbol{k}\right)$,
$g_{n\alpha\beta}\left(\boldsymbol{k}\right)=\left(\partial_{\alpha}\hat{\boldsymbol{d}}_{n}\cdot\partial_{\beta}\hat{\boldsymbol{d}}_{n}\right)/4$
and $\Omega_{n\alpha\beta}\left(\boldsymbol{k}\right)=\hat{\boldsymbol{d}}_{n}\cdot\left(\partial_{\alpha}\hat{\boldsymbol{d}}_{n}\times\partial_{\beta}\hat{\boldsymbol{d}}_{n}\right)/2$
respectively correspond to the quantum metric and Berry curvature
of electrons \citep{XiaoD2010RMP,LiuTY2024NSR}. The real and imaginary
part of $\mathcal{G}_{n\alpha\beta}\equiv D_{n\alpha\beta}-im_{n\alpha\beta}$
are associated with the band Drude weight and orbital magnetic moment
\citep{Kohn1964PR,Resta2018jpcm}. It is worth mentioning that the
imaginary part of $\mathcal{H}_{n\alpha\beta}\equiv D_{n\alpha\beta}^{E}-im_{n\alpha\beta}^{E}$
refers to the HMM of electrons \citep{QinT2011PRL}, while the real
part could be regarded as the generalized heat Drude weight, in analogy
to the band Drude weight in electric conductivity \citep{Shastry2006PRB}
(Details in the Supplemental Material \citep{SMs}). In fact, the
significance of the unusual quantum geometric tensor $\mathcal{H}_{n\alpha\beta}$
is rarely discussed before. 
\begin{table}
\begin{tabular}{|c|c|c|}
\hline 
 & Real & Imaginary\tabularnewline
\hline 
\hline 
$\mathcal{F}_{n\alpha\beta}$ & \begin{cellvarwidth}[t]
\centering
quantum metric

$g_{n\alpha\beta}$
\end{cellvarwidth} & \begin{cellvarwidth}[t]
\centering
Berry curvature

$\Omega_{n\alpha\beta}$
\end{cellvarwidth}\tabularnewline
\hline 
$\mathcal{G}_{n\alpha\beta}$ & \begin{cellvarwidth}[t]
\centering
band Drude weight

$D_{n\alpha\beta}$
\end{cellvarwidth} & \begin{cellvarwidth}[t]
\centering
orbital magnetic moment

$m_{n\alpha\beta}$
\end{cellvarwidth}\tabularnewline
\hline 
$\mathcal{H}_{n\alpha\beta}$ & \begin{cellvarwidth}[t]
\centering
heat Drude weight

$D_{n\alpha\beta}^{E}$
\end{cellvarwidth} & \begin{cellvarwidth}[t]
\centering
heat magnetic moment

$m_{n\alpha\beta}^{E}$
\end{cellvarwidth}\tabularnewline
\hline 
\end{tabular}

\caption{\label{sixQGQ} Sestet of quantum geometric quantities associated
with $\mathcal{F}_{n\alpha\beta}$, $\mathcal{G}_{n\alpha\beta}$ and
$\mathcal{H}_{n\alpha\beta}$ of Bloch electrons. The real and imaginary
parts of the quantum geometric tensors correspond to the symmetric
part and antisymmetric part, respectively.}
\end{table}

The electron-phonon coupling $\hat{H}_{e-ph}$ can be modeled by coupling
the momentum with the local frame field $e_{j}^{\mu}$ via $p_{j}\rightarrow p_{\mu}e_{j}^{\mu}$,
where $e_{j}^{\mu}=\delta_{j}^{\mu}-\partial u_{j}\left(\boldsymbol{r}\right)/\partial r_{\mu}$
describes the local stretching and rotation of lattice structure \citep{Barkeshli2012PRB,Hughes2011PRL,Shapourian2015PRB},
and summation over repeated indices is implied henceforth. We further
expand the factor $S_{nn^{\prime}}^{\alpha\beta}(\boldsymbol{k},\boldsymbol{q})$
with respect to $q$ to third order and have (Details are given in
the Supplemental Material \citep{SMs}) the intraband part 
\begin{align}
 & S_{n}^{\alpha\beta}(\boldsymbol{k},\boldsymbol{q})\nonumber \\
= & \frac{1}{2}\left(1+\frac{1}{2}\boldsymbol{q}\cdot\boldsymbol{\nabla}_{k}\right)\left[\left(\boldsymbol{q}\cdot\boldsymbol{k}\right)^{2}v_{n\boldsymbol{k}}^{\alpha}v_{n\boldsymbol{k}}^{\beta}\right]\nonumber \\
+ & \left(\boldsymbol{q}\cdot\boldsymbol{k}\right)^{2}iq_{\nu}\mathrm{Im}\left[v_{n\boldsymbol{k}}^{\alpha}\mathcal{G}_{n\nu\beta}\left(\boldsymbol{k}\right)+v_{n\boldsymbol{k}}^{\beta}\mathcal{G}_{n\alpha\nu}\left(\boldsymbol{k}\right)\right]\label{IntraSnn}
\end{align}
and the interband part
\begin{align}
 & S_{nm}^{\alpha\beta}(\boldsymbol{k},\boldsymbol{q})\nonumber \\
= & \left(1+\frac{1}{2}\boldsymbol{q}\cdot\boldsymbol{\nabla}_{k}\right)\left[\left(\boldsymbol{q}\cdot\boldsymbol{k}\right)^{2}\mathcal{H}_{n\boldsymbol{k}}^{\alpha\beta}\right]\nonumber \\
+ & \left(\boldsymbol{q}\cdot\boldsymbol{k}\right)^{2}q_{\nu}\left[v_{nm\boldsymbol{k}}^{\alpha}\mathcal{G}_{n\nu\beta}\left(\boldsymbol{k}\right)+v_{mn\boldsymbol{k}}^{\beta}\mathcal{G}_{n\alpha\nu}\left(\boldsymbol{k}\right)\right]\label{InterSnm}
\end{align}
where $v_{mn\boldsymbol{k}}^{\alpha}\equiv\frac{1}{2}\left(v_{m\boldsymbol{k}}^{\alpha}+v_{n\boldsymbol{k}}^{\alpha}\right)$
is the mean velocity of two bands and will disappear for the particle-hole
symmetric bands. Note that summation over repeated indices is assumed.
According to the definition of $\gamma$ matrix in Eq. $\left(\ref{gammaM}\right)$,
the linear phonon dichroism in Eq. (\ref{gamlt}) can be written as
the leading-order contribution (up to $q^{2}$)
\begin{align}
\gamma^{\bar{A}} & =\frac{-\hbar}{2\omega\rho}\int\frac{d^{2}\boldsymbol{k}}{(2\pi)^{2}}\left(\boldsymbol{q}\cdot\boldsymbol{k}\right)^{2}\left\{ \frac{1}{2}\sum_{n}f_{n,n}\left(\boldsymbol{k},\boldsymbol{k}+\boldsymbol{q}\right)\right.\nonumber \\
 & \times v_{n\boldsymbol{k}}^{x}v_{n\boldsymbol{k}}^{y}\delta\left[\varepsilon_{nn}\left(\boldsymbol{k}+\boldsymbol{q},\boldsymbol{k}\right)-\hbar\omega\right]+\sum_{n\neq n'}D_{nxy}^{E}\left(\boldsymbol{k}\right)\nonumber \\
 & \times\left.f_{n,n^{\prime}}\left(\boldsymbol{k},\boldsymbol{k}+\boldsymbol{q}\right)\delta\left[\varepsilon_{n^{\prime}n}\left(\boldsymbol{k}+\boldsymbol{q},\boldsymbol{k}\right)-\hbar\omega\right]\right\} 
\end{align}
and the diagonal element, 
\begin{align}
\gamma_{\alpha\alpha} & =\frac{-\hbar}{2\omega\rho}\int\frac{d^{2}\boldsymbol{k}}{(2\pi)^{2}}\left(\boldsymbol{q}\cdot\boldsymbol{k}\right)^{2}\left\{ \frac{1}{2}\sum_{n}f_{n,n}\left(\boldsymbol{k},\boldsymbol{k}+\boldsymbol{q}\right)\right.\nonumber \\
 & \times\left(v_{n\boldsymbol{k}}^{\alpha}\right)^{2}\delta\left[\varepsilon_{nn}\left(\boldsymbol{k}+\boldsymbol{q},\boldsymbol{k}\right)-\hbar\omega\right]+\sum_{n\neq n'}D_{n\alpha\alpha}^{E}\left(\boldsymbol{k}\right)\nonumber \\
 & \times\left.f_{n,n^{\prime}}\left(\boldsymbol{k},\boldsymbol{k}+\boldsymbol{q}\right)\delta\left[\varepsilon_{n^{\prime}n}\left(\boldsymbol{k}+\boldsymbol{q},\boldsymbol{k}\right)-\hbar\omega\right]\right\} 
\end{align}
where $D_{nxy}^{E}\left(\boldsymbol{k}\right)$ is the heat Drude
weight of electrons in the $n$th band and can be regarded as the
cousin of the quantum metric with $\alpha=x,y$. It can be seen that
the linear phonon dichroism is controlled by the quantum metric (See
the full expressions of $\gamma^{\bar{A}}$ and $\gamma_{\alpha\alpha}$
\citep{SMs}). 
\begin{figure}
\includegraphics[scale=0.55]{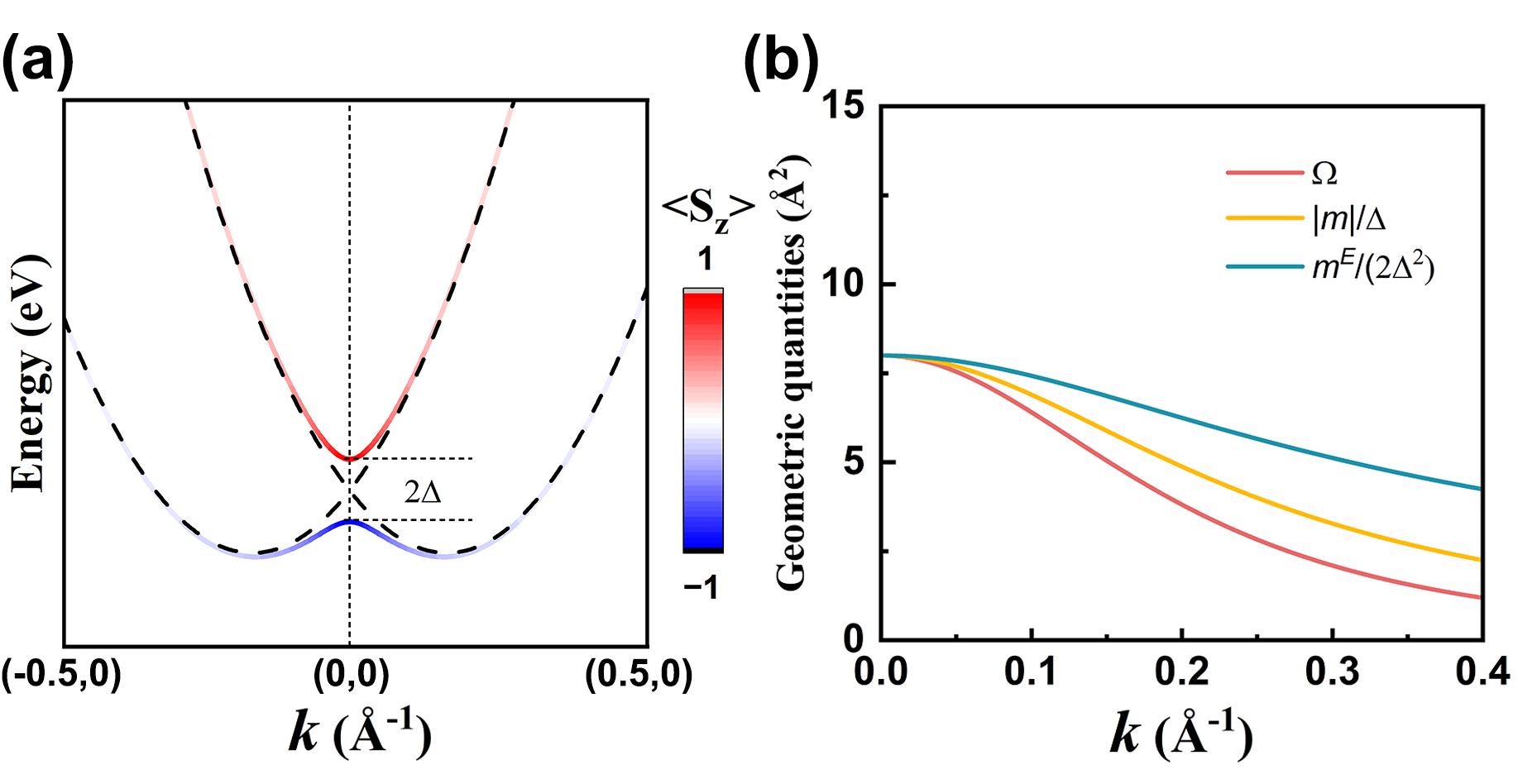}\caption{(a) The energy band of Rashba 2DEG in the presence (colored solid
lines) or absence (dashed line) of out-of-plane Zeeman fields. (b)
The corresponding radial distribution of the three quantum geometric
quantities: Berry curvature, orbital magnetic moment and HMM of the
$E_{-}$ band. \label{fig2}}
\end{figure}

Let us turn to the circular phonon dichroism. Remarkably, the leading-order
term quadratic in $q$ of $\mathrm{Im}\left[S_{nm}^{\alpha\beta}\right]$
is directly related to the HMM of electrons from $\mathcal{H}_{n\boldsymbol{k}}^{\alpha\beta}$
in Eq. $\left(\ref{InterSnm}\right)$ and leads to a basic and useful
relation of the circular phonon dichroism
\begin{alignat}{1}
\gamma^{A}\left(\bm{q},\omega\right) & \approx-\frac{\pi\hbar}{4\omega\rho}\sum_{n\neq n'}\int\frac{d^{2}\boldsymbol{k}}{(2\pi)^{2}}\left(\bm{q}\cdot\bm{k}\right)^{2}m_{n\bm{k}}^{E}\nonumber \\
 & \times f_{n,n^{\prime}}\left(\boldsymbol{k},\boldsymbol{k}+\boldsymbol{q}\right)\delta\left[\varepsilon_{n^{\prime}n}\left(\boldsymbol{k}+\boldsymbol{q},\boldsymbol{k}\right)-\hbar\omega\right],\label{Gamqq}
\end{alignat}
where $m_{n\bm{k}}^{E}$ is the HMM of electrons and the factor of
$q^{2}/\omega\rho$ corresponds to the square of strength of electron-phonon
interaction \citep{Kittel1987qts}. 

In order to gain the fundamental quantities of electronic states,
it is quite instructive to investigate the sum rule \citep{pines1966}.
In analogy to the well-known $f$-sum rule for the particle density,
we consider a proper integral of $\gamma^{A}\left(\bm{q},\omega\right)\omega$
over $\omega$ as $I_{\gamma^{A}}\equiv\int\omega\gamma^{A}\left(\bm{q},\omega\right)d\omega$,
\begin{align}
 & I_{\gamma^{A}}\nonumber \\
\approx & -\frac{\pi\hbar}{4\rho}q_{\alpha\beta}\sum_{n\neq n'}\int\frac{d^{2}\boldsymbol{k}}{(2\pi)^{2}}k_{\alpha}k_{\beta}m_{n\bm{k}}^{E}f_{n,n^{\prime}}\left(\boldsymbol{k},\boldsymbol{k}+\boldsymbol{q}\right).\label{Gamqq-omm}
\end{align}
with $q_{\alpha\beta}\equiv\left(q_{x}^{2},2q_{x}q_{y},q_{y}^{2}\right)$.
In order to unravel the clean physical picture, we consider the particle-hole
symmetric bands at low temperatures \citep{origin1}. For simplicity,
we set $\bm{q}=\left(q_{x},q_{y}\right)$ is small relative to the
Fermi vector $k_{F}$ and the Fermi level lies in the valence band.
Thus, the integration $I_{\gamma^{A}}$ could be further simplified and becomes 
\begin{equation}
I_{\gamma^{A}}\approx-\frac{\pi\hbar}{4\rho}q_{\alpha\beta}\sum_{n}\int\frac{d^{2}\boldsymbol{k}}{(2\pi)^{2}}k_{\alpha}k_{\beta}f_{n}\left(\boldsymbol{k}\right)m_{n\bm{k}}^{E}\label{Intgam2}
\end{equation}
The right hand side of Eq. (\ref{Intgam2}) is the sum of quadrupole
of $m_{n\bm{k}}^{E}$ of filled states. It shows clearly that the
circular phonon dichroism is primarily determined by $m_{n\bm{k}}^{E}$
of electrons. This formula provides a guiding principle to detect
the HMM of electrons in experiments \citep{DetectHMM}, similar to
the magnetic circular dichroism of light or X-ray for the orbital
magnetization in magnetic materials \citep{VANDERLAAN2014CCR}. Eqs.~(\ref{Gamqq})-(\ref{Intgam2})
are central results of the present work. Next we would demonstrate
the significance of quantum metric and HMM in phonon dichroisms in
a typical quantum material. 
\begin{figure}[tp]
\includegraphics[scale=0.25]{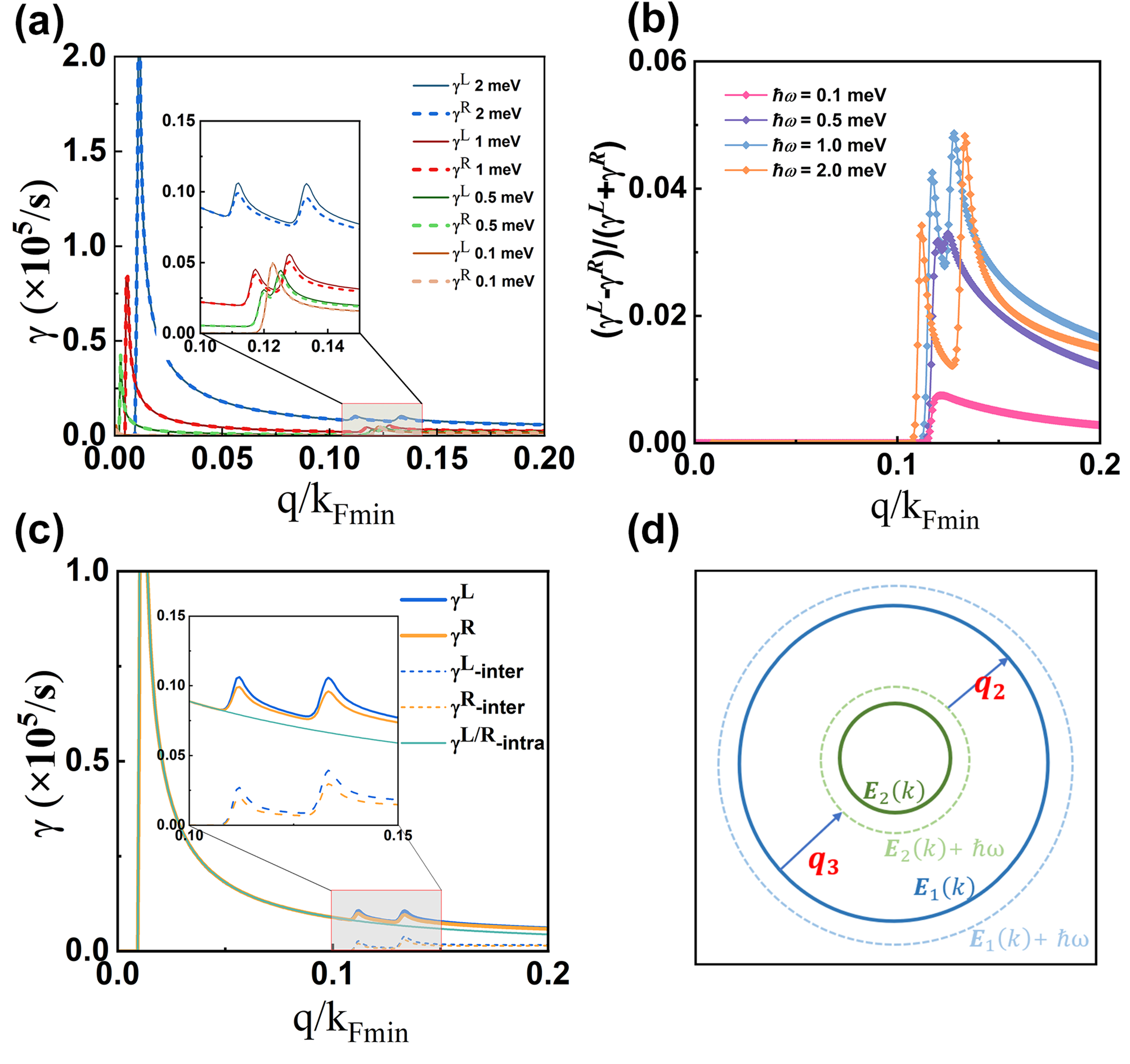}

\caption{\label{fig3}(a)-(d) Damping coefficients of $\gamma^{R/L}$ phonon
modes versus the phonon wave-vector $\boldsymbol{q}$ of Rashba 2DEG.
(a) $\boldsymbol{q}-$dependence of $\gamma^{R/L}$ at several typical
energies and $k_{Fmin}=\sqrt{\left(\mu-\Delta\right)/t}$. Insert
shows the region for evident difference and $\gamma^{R}$ and $\gamma^{L}$.
(b) Two-peak signature in ratio of $\left(\gamma^{L}-\gamma^{R}\right)/\left(\gamma^{R}+\gamma^{L}\right)$
at phonon energies of $\text{\ensuremath{\hbar}}\omega=0.5,\:1.0\:\mathrm{meV}$.
(c) Decomposition of strength of circular phonon dichroism in terms
of the interband and intraband processes. (d) Origin of two-peak structure
in the damping coefficients of $\gamma^{R/L}$. We fixed $\boldsymbol{q}$
at a direction $\arctan(q_{y}/q_{x})=\pi/4$ and $q>0$. $\mu=0.1\:\mathrm{eV}$
is relative to the band bottom. }
\end{figure}

\textit{\textcolor{black}{Rashba 2DEG.-{}-}}We first consider the
generic 2DEG with Rashba SOC in the presence of an out-of-plane Zeeman
field that can well describe the electrons at the surface of metals
\citep{BychkovJETP1984}, $\mathrm{LaAlO_{3}/SrTiO_{3}}$ oxide heterointerfaces
\citep{Ohtomo2004Nature} and $\mathrm{KTaO_{3}}$-based heterostructures
\citep{LiuCJ2021Science}. It acts as a model system for investigation
of low-dissipation electromagnetic responses \citep{Culcer2003PRB,ZhouJH2015PRB,Hamdi2023NP},
chiral spin textures \citep{LiXP2014PRL} and topological superconductivity
\citep{Gorkov2001PRL}. The model is given as
\begin{equation}
H(\boldsymbol{k})=tk^{2}+\Delta\sigma_{z}+\lambda(k_{x}\sigma_{y}-k_{y}\sigma_{x}),\label{HRashba2deg}
\end{equation}
where $t$ breaks particle-hole symmetry and $\Delta$ breaks time-reversal
symmetries and gives rise to a band gap $2\Delta$ at the $\Gamma$
point (Fig. \ref{fig2}(a)) and $k=\sqrt{k_{x}^{2}+k_{y}^{2}}$. Importantly,
the Rashba SOC $\lambda$ breaks the spatial inversion symmetry and
is vital to the emergence of nontrivial quantum geometry tensors in
momentum space. Some manipulations lead to the Berry curvature of
the two bands ($E_{\pm}=tk^{2}\pm\sqrt{\Delta^{2}+\lambda^{2}k^{2}}$),
$\Omega_{\pm}^{xy}\left(\boldsymbol{k}\right)=\frac{\mp\lambda^{2}\Delta}{2\left(\Delta^{2}+k^{2}\lambda^{2}\right)^{3/2}}$
and the corresponding orbital/heat magnetic moments $m_{\pm}\left(\boldsymbol{k}\right)=\frac{-\lambda^{2}\Delta}{2\left(\Delta^{2}+k^{2}\lambda^{2}\right)}$
and $m_{\pm}^{E}\left(\boldsymbol{k}\right)=\frac{\pm\lambda^{2}\Delta}{\left(\Delta^{2}+k^{2}\lambda^{2}\right)^{1/2}}$
(The real parts of quantum geometric tensors are given in Supplemental
Material \citep{SMs}). As shown in Fig. \ref{fig2}(b), $m_{n}^{E}\left(\boldsymbol{k}\right)$
exhibits a smooth distribution in momentum space than $m_{\pm}\left(\boldsymbol{k}\right)$
and $\Omega_{n}\left(\boldsymbol{k}\right)$. In the strong-exchange
field limit ($\Delta^{2}\gg k^{2}\lambda^{2}$), the orbital/heat
magnetic moments reduce to be $m_{\pm}\left(\boldsymbol{k}\right)\approx-\frac{\lambda^{2}}{2\Delta}\left(1-k^{2}\lambda^{2}/\Delta^{2}\right)$
and $m_{\pm}^{E}\left(\boldsymbol{k}\right)\approx\pm\lambda^{2}\left(1-k^{2}\lambda^{2}/2\Delta^{2}\right)$.
Clearly, the Rashba SOC plays a primary role in the circular phonon
dichroisms. 
\begin{figure}
\includegraphics[scale=0.6]{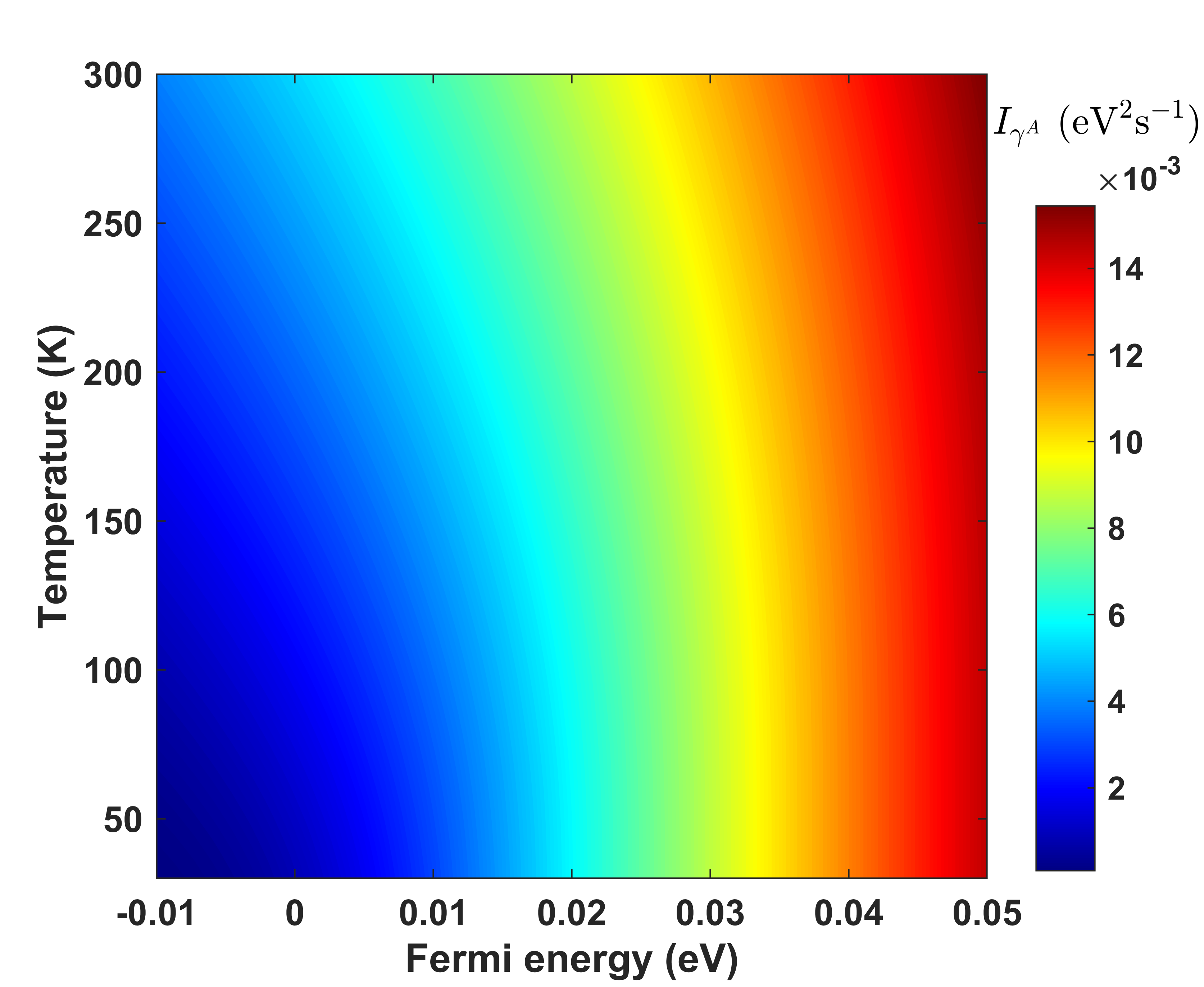}

\caption{\label{fig4} Temperature- and Fermi energy- dependence of the integration
of the circular phonon dichroism of Rashba 2DEG with $q/k_{Fmin}=0.04$.}
\end{figure}

With the aid of Eq. $\left(\ref{gammaM}\right)$, we could calculate
the circular phonon dichroism when $\boldsymbol{q}$ is along $\arctan(q_{y}/q_{x})=\pi/4$
direction and depict the results in Fig.$\:$\ref{fig3}. There are
several salient features in $\gamma^{R/L}$. First, in the long-wavelength
region (small $q$), the absorptions of the right- and left-hand circularly
polarized phonons are strong but almost identical, implying a vanishing
circular phonon dichroism. Second, as the phonon wave-vector increases,
the difference of absorption coefficients between oppositely circularly
polarized phonons becomes pronounced at $q/k_{Fmin}\sim1.2$ and forms
a two-peak structure as shown in Figs.$\:$\ref{fig3}(a)-(c). Third,
Fig.$\:$\ref{fig3}(d) shows the interband electron-hole pair excitations
between states with opposite spins induced by phonons (labeled by
$\boldsymbol{q}_{2}$ and $\boldsymbol{q}_{3}$) are dominant in $\gamma^{L}-\gamma^{R}$,
similar to the 3D Weyl semimetals \citep{LiuDH2017PRL}. The intraband
contribution instead causes a global change of the absorption coefficients.
Furthermore, we find the double peaks mainly originate from the asymmetry
of interband electron-hole pair excitations in the two regions near
the Fermi surface. In addition, we could evaluate the related circular
phonon dichroism and have $\left(\gamma^{L}-\gamma^{R}\right)/\bar{c}\sim10\mathrm{/m}$,
where $\bar{c}=\text{\ensuremath{\left(c_{l}+c_{t}\right)}}/2\approx10^{3}\mathrm{m}/\mathrm{s}$
is the average sound velocity. Noted that, in our numerical calculations,
we focus on the acoustic phonons and choose $c_{l/t}$ as the order
of $10^{3}\mathrm{m}/\mathrm{s}$ throughout this work. 

We also calculate the $f$-sum rule integration of the circular phonon
dichroism in Eq. (\ref{Gamqq-omm}) of the Rashba 2DEG and depict
the dependence of temperature and Fermi energy of $I_{\gamma^{A}}$
in Fig. \ref{fig4}. It is clear that, as temperature or the Fermi
level increases, the circular phonon dichroism gets greatly stronger
due to the enhancement of population of phonon induced electron-hole excitations. 

Let us briefly discuss the experimental detection of phonon dichroisms.
The difference of absorption between the left and right circularly
polarized phonons can be accessed by the pulse-echo ultrasound technique
\citep{Luthi2006PhysAI}, whereas the phonon polarization of linearly
polarized acoustic waves can be detected by Raman spectroscopy \citep{Faugeras2011PRL}.
Recently, interface-specific nonlinear optical technique has been
used to probe the polaronic signature in $\mathrm{LaAlO_{3}/SrTiO_{3}}$
interface \citep{LiuXY2023SA}, providing a feasible platform to detect
the quantum-geometry modified phonon dichroism therein. Very recently,
the circular dichroism of phonon has been reported in a magnetic Weyl
semimetal \citep{YangRun2025PRL,Che2025PRL} and makes the present
predictions be tested experimentally. It should be emphasized that
together with the spin- and angle-resolved circular dichroism photoemission
spectroscopy \citep{Kang2025NP}, the phonon dichroisms allow us to
extract the complete information of all of six quantum geometric quantities
(in Table. \ref{sixQGQ}) of electrons in solids.

\textit{\textcolor{black}{Conclusions.-{}-}}We established the deep
connection between phonon dichroisms and the unusual quantum geometry
of electron bands in noncentrosymmetric magnetic metals. Notably,
we found that the circular phonon dichroism is directly related to
the HMM of electrons through the sum rule. The key results are examined
in the typical Rashba 2DEG and can be accessible by various spectroscopy
techniques. Our findings provide a novel contact-free acoustic method
for detecting the \textcolor{black}{fundamental quantum geometry}
in particular the HMM of electrons. It would inspire the investigation
of the quantum geometry effect on the fascinating quantum phenomena
in other coupled systems among electron and bosonic excitations in
a large variety of quantum materials. 

\textcolor{black}{The authors thank D.-H. Liu, W.-Y. Shan, J.-R.
Shi and Y.-H. Zhang for useful discussions. }This work was financially supported by
the National Key R\&D Program of the MOST of China (Grant No. 2024YFA1611300),
the National Natural Science Foundation of China (Grant No. 12574059
and No. 12174394), HFIPS Director's Fund (Grant No. BJPY2023B05),
Anhui Provincial Major S\&T Project (s202305a12020005) and the Basic
Research Program of the Chinese Academy of Sciences Based on Major
Scientific Infrastructures (Grant No. JZHKYPT-2021-08) and the High
Magnetic Field Laboratory of Anhui Province under Contract No. AHHM-FX-2020-02.

\bibliographystyle{apsrev4-1}
\bibliography{PDQGT}

\begin{thebibliography}{66}
\bibitem [{ori()}]{origin1}%
  \BibitemOpen
  \href@noop {} {\bibinfo  {journal} {Since the acoustic phonon energy is
  typically much smaller than the Fermi energy, the initial and final states of
  the electron should be close to the Fermi surface during phonon absorption.
  Further taking into account the spin-orbit coupling, the spin-momentum
  locking of electrons could affect the phonon behaviors through the
  electron-phonon interaction}\ }\BibitemShut {NoStop}%
\bibitem [{Det()}]{DetectHMM}%
  \BibitemOpen
\bibfield  {journal} {  }\href@noop {} {\bibinfo  {journal} {The generalization
  of the sum rule and a procedure to extract the electron heat and orbital
  magnetic moments from the experimental measurements on phonon dichroism are
  also discussed~\cite{SMs}}\ }\BibitemShut {NoStop}%
\end{thebibliography}

\end{document}